\documentclass[3p,times,nopreprintline]{elsarticle}

\usepackage{lineno}

\usepackage{microtype}
\usepackage{hyperref}
\biboptions{sort&compress}
\usepackage[dvipsnames]{xcolor}
\usepackage[inline]{enumitem}
\usepackage{amsmath}
\usepackage{amssymb}
\usepackage[T1]{fontenc}
\usepackage{newtxtext,newtxmath}
\usepackage{cleveref}
\usepackage{booktabs}
\usepackage{adjustbox}

\hypersetup{
    colorlinks,
    linkcolor={red!50!black},
    citecolor={blue!50!black},
    urlcolor={blue!80!black}
}

\setlist{label=(\textit{\roman*})}

\newcommand{\hidetext}[1]{}

\begin{document}

\begin{frontmatter}




\title{Progress in ${\cal CP}$ violating top-Higgs coupling at the LHC with Machine Learning}


\author[a]{A. Hammad}
\author[b]{Adil Jueid}

\address[a]{Theory Center, IPNS, KEK,  1-1 Oho, Tsukuba, Ibaraki 305-0801, Japan}
\address[b]{Particle Theory and Cosmology Group, Center for Theoretical Physics of the Universe, \\ Institute for Basic Science (IBS), 34126 Daejeon, Republic of Korea}

\begin{abstract}
A precise measurement of the top-Higgs coupling is essential in particle physics, as it offers a powerful probe of potential new physics beyond the Standard Model (BSM), particularly scenarios involving ${\cal CP}$ violation, which is a key condition in addressing the problem of baryon asymmetry of the universe. In this article, we review the recent progress in the studies of the the top-Higgs coupling at the Large Hadron Collider (LHC). We briefly highlight the recent Machine Learning (ML) algorithms being used and their role in constraining the ${\cal CP}$ phase of the top-Higgs coupling with an emphasis on the future potential of beyond-the-traditional methods such as transformers and heterogeneous graphs in these studies.
\end{abstract}


\end{frontmatter}

\tableofcontents


\section{Introduction}
\label{sec:intro}

The mechanism behind ${\cal CP}$ violation in the quark sector remains one of the most intriguing puzzles of the Standard Model (SM). Since its first discovery in $K$-meson decays in 1964 \cite{Christenson:1964fg}, ${\cal CP}$ violation has been an active area of research across both experimental and theoretical particle-physics communities. Studies of ${\cal CP}$ violation in both collider and table-top experiments are crucial, as it constitutes one of the three necessary conditions for baryogenesis originally proposed by Sakharov \cite{Sakharov:1967dj}. Measurements of various asymmetries in the $D$- and $B$-meson sectors strongly indicate that ${\cal CP}$ violation can be solely explained by the Kobayashi-Maskawa (KM) mechanism \cite{Kobayashi:1973fv}. It is found that ${\cal CP}$ violation generated by KM mechanism is not enough to generate the desired antimatter asymmetry in the universe even if the phase transition were first order \cite{Gavela:1993ts,Huet:1994jb,Gavela:1994dt}. Many extensions of the SM can accommodate for additional sources of ${\cal CP}$ violation either through explicit or spontaneous breaking terms. However, these models have been strongly constrained by very precise measurements of permanent electric dipole moments (EDMs) of various systems \cite{Andreev:2018ayy,Cairncross:2017fip,Abel:2020gbr} with sensitivities that are many orders of magnitude above the state-of-art precise predictions in the SM  \cite{Yamaguchi:2020eub,Yamaguchi:2020dsy}. Despite their significant sensitivity, EDMs are by definition inclusive observables, making them unlikely to probe the underlying nature of ${\cal CP}$ violation. In contrast, collider observables being exclusive in their nature are {\it apriori} able to pin down the origin of ${\cal CP}$ violation despite their weaker sensitivities as compared to EDM observables.  The interest to build collider observables to search for ${\cal CP}$ in processes involving the Higgs boson has grown specifically after the discovery of the SM Higgs boson at the Large Hadron Collider (LHC) in July 2012 \cite{Aad:2012tfa,Chatrchyan:2012xdj}. In particular, constraining the spin and ${\cal CP}$ properties of the newly discovered Higgs boson plays a central role in various global statistical analyses \cite{Ellis:2012hz,Ellis:2013lra,Belanger:2013xza,Dumont:2014wha,Corbett:2015ksa}. Studies of the ${\cal CP}$ properties at high-energy collider experiments have been performed in the last two decades especially exploring the potential probe the Higgs couplings to fermions and gauge bosons at the LHC  \cite{Schmidt:1992et,Berge:2008dr,Shu:2013uua,Bishara:2013vya,Harnik:2013aja,Ellis:2013yxa,Yue:2014tya,Chang:2014rfa,Englert:2014pja,Berge:2014sra,He:2014xla,Boudjema:2015nda,Askew:2015mda,Hagiwara:2016zqz,Ferreira:2016jea,Jozefowicz:2016kvz,Mileo:2016mxg,Brehmer:2017lrt,AmorDosSantos:2017ayi,Azevedo:2017qiz,Bernlochner:2018opw,Cirigliano:2019vfc,Faroughy:2019ird,Ren:2019xhp,Cao:2020hhb,Bahl:2020wee,Cheung:2020ugr,Azevedo:2020fdl,Azevedo:2020vfw,Martini:2021uey,Wu:2021xsj,Bahl:2021dnc,Barman:2021yfh,Yu:2022kcj,Bahl:2022yrs,Bahl:2023qwk,Ackerschott:2023nax,Miralles:2024huv,Thomas:2024dwd,Esmail:2024gdc,Esmail:2024jdg,Bahl:2024tjy} and future colliders \cite{Bower:2002zx,Desch:2003mw,Desch:2003rw,Asakawa:2003dh,BhupalDev:2007ftb,Berge:2013jra,He:2014xla,Bernreuther:2017cyi,Hagiwara:2017ban,Khanpour:2017cfq,Ma:2018ott,Karadeniz:2019upm,Cassidy:2023lwd}. Among these phenomenological studies, a particular interest was given to the analysis of ${\cal CP}$ violation in the top-Higgs interaction for various reasons. First, this coupling being the largest among all the SM Higgs couplings influence indirectly the decay partial widths of the SM Higgs boson into $\gamma\gamma, gg$ and $\gamma Z$ as well as the cross sections of single and pair production of the Higgs boson. Second, unlike the ${\cal CP}$ violation in the other Higgs coupling, the top-Higgs coupling is has the highest sensitivity to the nature of electroweak symmetry breaking as it can be affected by the presence of extra Higgs states with masses around $300$--$700$ GeV.

The determination of the ${\cal CP}$ properties of the top-Higgs interaction relies on crucial factors including the suppression of the large associated backgrounds at the LHC which lead to the definition of the so-called signal regions (SRs), {\it i.e.} kinematical regions that are more ``signal" enriched. This is followed by a measurement of the ${\cal CP}$-violating observables which are then used to constrain the phase of the top-Higgs coupling. Early methods relies on the cut-and-count methods which are done by sequential selection cuts on key kinematical variables. However, these methods being sequential does not provide {\it efficient} definition of the SRs and may even lead to a significant loss of the sensitivity of these ${\cal CP}$-odd observables. A basic approach to address this issue is to use Machine Learning (ML) algorithms to define ``clean'' SRs and even extract information on the ${\cal CP}$ phase by combining the basic high- and low-level variables with ${\cal CP}$-odd variables in a post-fit regime. At the foundation lies the Boosted Decision Trees (BDTs) — the first machine learning method applied to collider analyses (see {\it e.g.} Refs. \cite{Choudhury:2024crp,Cheng:2017rdo,Dutta:2023jbz,Cornell:2021gut,Antusch:2019eiz,Antusch:2020vul,Antusch:2020ngh,Antusch:2018bgr,DelleRose:2018ndz}). BDTs are powerful classifiers at the LHC, effectively isolating the signal from backgrounds using high-level variables and jet substructure observables. Their ability to capture the non-linear correlations enhances event selection in new physics searches. By combining many weak learners, BDTs improve sensitivity while maintaining robustness to systematic uncertainties. However, they are prone to overfitting and especially in case with too many or very deep trees. Moreover, the performance of BDTs relies heavily on feature engineering, requiring careful preselection of input events. One of the most effective test statistics for classification is the event likelihood ratio at the parton level. Yet, this ratio becomes intractable when real data are used due to latent effects like showering, hadronization, and detector response. ML offers a solution by framing likelihood estimation as an inference problem. With an appropriate choice of the loss function, ML models can approximate the event-level likelihood ratio -- a strategy first implemented in the \textsf{MadMiner} toolkit \cite{Brehmer:2019xox,MadMiner_code}. An alternative approach is the use of parameterized neural networks, introduced in collider studies in Refs. \cite{Cranmer:2015bka,Chen:2023ind,Baldi:2016fzo,Chen:2020mev,Esmail:2024gdc}. These networks are trained on benchmark points with different ${\cal CP}$ phase values, enabling them to interpolate smoothly across the entire CP phase range at inference time. This method is architecture-agnostic and can be applied to standard Multi-Layer Perceptrons (MLPs) as well as more advanced models like Graph Neural Networks (GNNs). 

In this article, we review the recent machine learning methods for constraining the ${\cal CP}$ phase in the top-Higgs interaction channel and discuss future directions to improve current analyses. In section \ref{sec:physics} we discuss the basic physical observables in constraining the ${\cal CP}$ phase in the top-Higgs interactions and their first use in LHC analyses. A discussion of the first machine learning methods being used in experimental analyses and phenomenological sensitivity projections will be explored in section \ref{sec:new}. We first give a brief overview of the various methods being used for studying $t\bar{t}H$ coupling which include BDTs, DL-based likelihood methods, MLPs and GNNs. Then we close section \ref{sec:new} with a summary of the  future sensitivities. Section \ref{sec:prospects} will be devoted to a discussion of the state-of-art methods such as those employing transformers and heterogeneous graphs and how they can be used to the probes of ${\cal CP}$ violation in the top-Higgs sector. We draw our conclusions in section \ref{sec:summary}.

\begin{figure}[!t]
    \centering
    \includegraphics[width=0.32\linewidth]{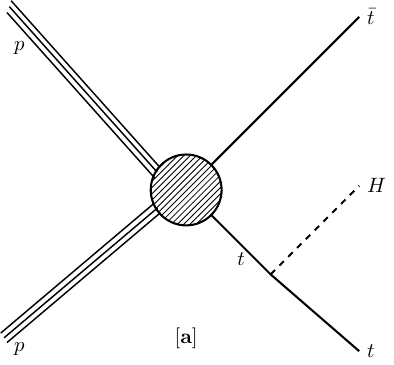}
    \includegraphics[width=0.32\linewidth]{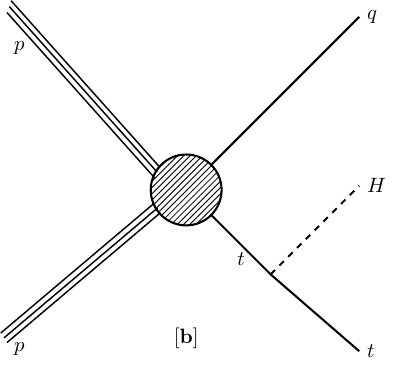}
    \includegraphics[width=0.32\linewidth]{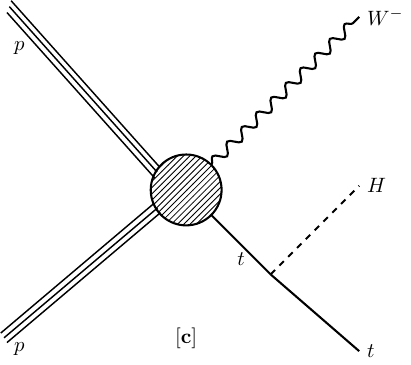}
    \caption{Diagrams for the associated top-Higgs production at the LHC. We show the production of the Higgs boson in association with $t\bar{t}$ pair (left), single production in association with $t\bar{q}$ (middle) and single production production with association $tW^-$ (right). The Feynman diagrams were made using \textsf{TickZ-Feynman} \cite{Ellis:2016jkw}.}
    \label{fig:tH:processes}
\end{figure}

\section{Top-Higgs interaction: kinematics and dynamics}
\label{sec:physics}

\subsection{Top-Higgs associated production at the LHC}

\begin{figure}[!t]
    \centering
    \includegraphics[width=0.49\linewidth]{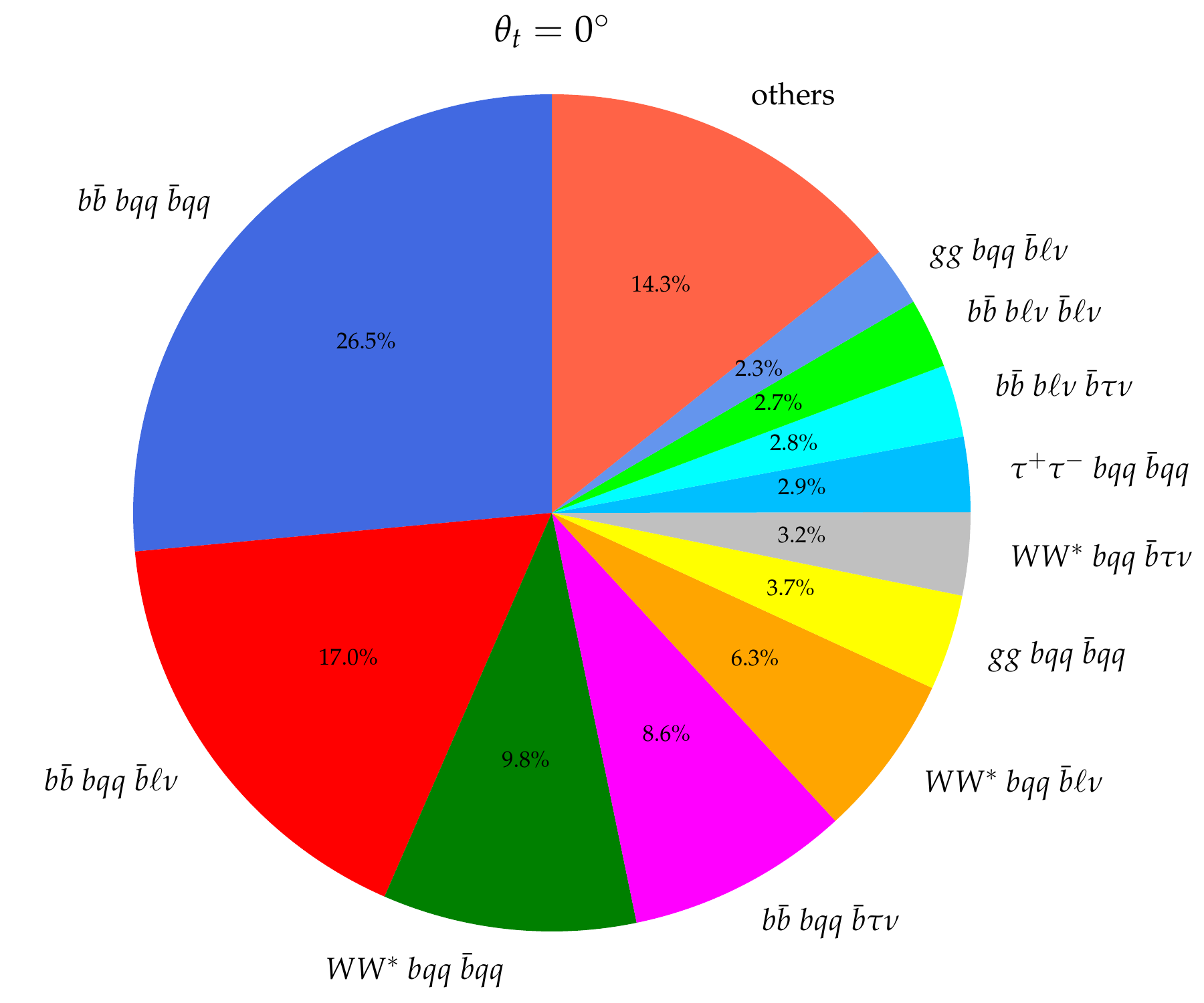}
    \hfill
    \includegraphics[width=0.49\linewidth]{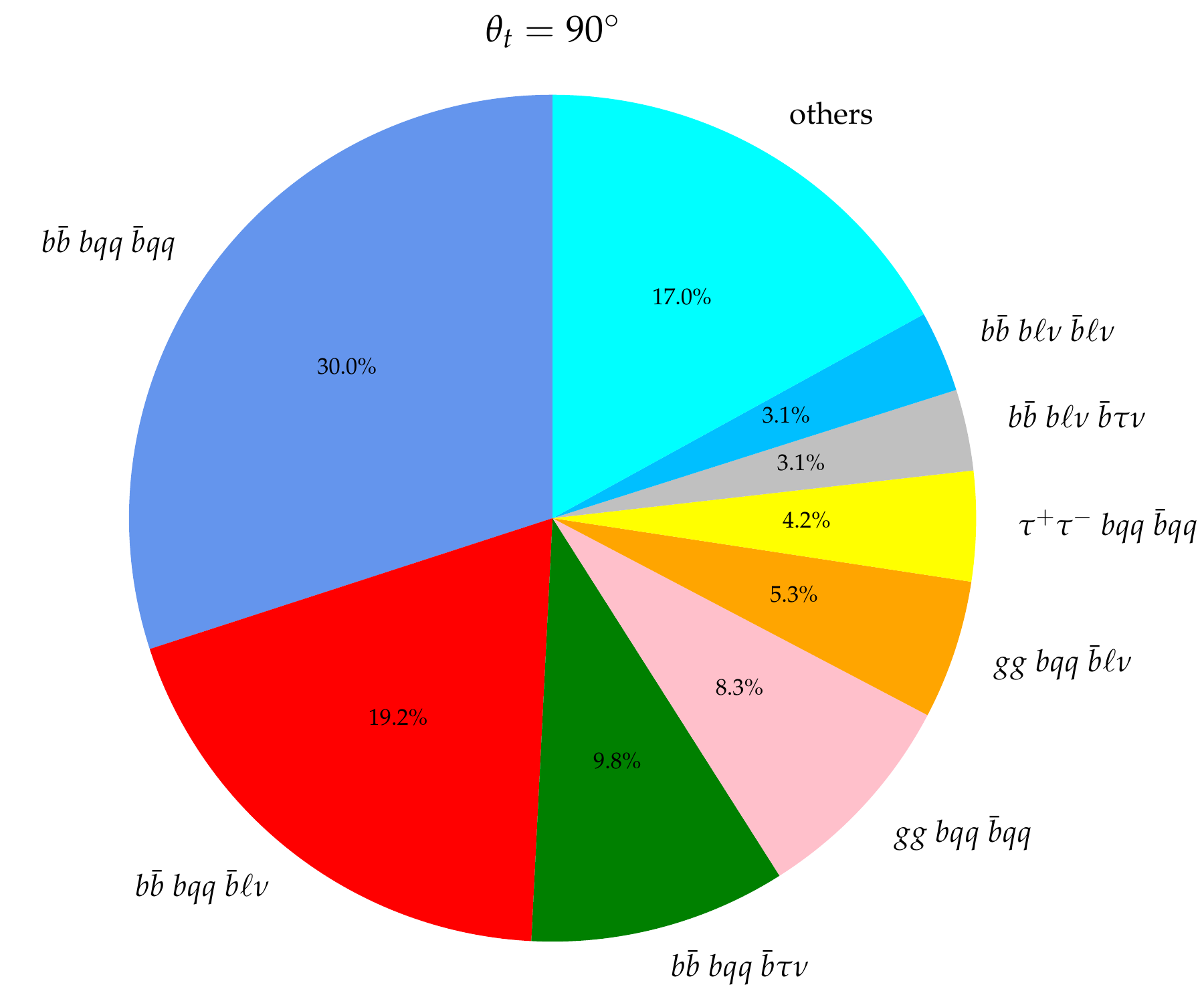}
    \caption{Decay branching ratios of $t\bar{t}H$ in the case of ${\cal CP}$ even (left panel) and ${\cal CP}$ odd (right panel). The label ``others" contains all the decays whose BRs are smaller than $2\%$.}
    \label{fig:decays}
\end{figure}

Including only terms up to dimension four, the generic Lagrangian for the top-Higgs interaction can be written as 
\begin{equation}
-\mathcal{L} \propto \frac{m_t}{\upsilon} \kappa_t \bar{t}\left(\cos\theta_t + \textrm{i} \sin\theta_t \gamma_5\right)t H,
\end{equation}
with $\upsilon \approx 246$ GeV being the vacuum expectation value (VEV) of the SM Higgs doublet, $\theta_t$ and $\kappa_t$ are two extra parameters. Note that a pure SM configuration is realized for $\kappa_t = 1$ and $\theta_t = 0^\circ$. The studies of the top-Higgs coupling in the direct channels are very promising given their uniqueness in both constraining both $\theta_t$ and $\kappa_t$ at the same time. At the LHC, the Higgs boson can be produced in association with the top quark through three different channels: $t\bar{t}H$ (Fig.  \ref{fig:tH:processes}a), $tqH$ (Fig. \ref{fig:tH:processes}b) and $tW^-H$ (Fig. \ref{fig:tH:processes}c). The first process is particularly interesting for two reasons: (i) it occurs with a higher rate compared to the other two processes\footnote{Single top production in association with a Higgs boson occurs with a smaller cross section as compared to the pair production mode. At NLO in QCD, the corresponding cross sections in the SM are 74.3 fb for $tHq$ and about $15.2$ fb for $tHW$ which are a factor of $\approx 7$ and $\approx 32$ smaller than that of the pair production mode (see Refs. \cite{Demartin:2015uha,Demartin:2016axk} for more details).} and (ii) its rate is directly proportional to $\kappa_t^2$. For $\theta_t=0^\circ$, we show below the scaling of the cross sections for these three processes (see Table 6 of Ref. \cite{ATLAS:2021vrm}):
\begin{eqnarray}
    \sigma(pp \to t\bar{t}H) &\propto& |\kappa_t|^2,  \nonumber \\
    \sigma(pp \to tqH) &\propto& 2.63 ~|\kappa_t|^2 + 3.58 ~|\kappa_W|^2 - 5.21 ~\kappa_W \kappa_t, \\
    \sigma(pp \to tW^-H) &\propto& 2.41 ~|\kappa_t|^2 + 2.31 ~|\kappa_W|^2 - 4.22 ~\kappa_W \kappa_t, \nonumber
\end{eqnarray}
where $\kappa_W$ is the coupling modifier of the Higgs-$W$ coupling defined as $\kappa_W^2 = \Gamma(H\to WW^*)/\Gamma(H\to WW^*)_{\rm SM}$. The single production of $H$t is interesting on its own due to its ability to probe the sign of $\kappa_t$. In what follows we discuss mostly the sensitivity on $\kappa_t$ and $\theta_t$ in the $t\bar{t}H$ channel. The cross section for $t\bar{t}H$ at the LHC is calculated up to NLO+NLL in QCD including electroweak corrections and parton-shower effects \cite{Beenakker:2001rj,Frederix:2011zi,Frixione:2014qaa}. For $\mu_0 = m_t + m_H/2$ and \textsf{PDF4LHC15} PDF sets \cite{Butterworth:2015oua}, the theoretical cross section at NLO+NLL in QCD is $\sigma(t\bar{t} H) = 509^{+21}_{-33}$ fb \cite{LHCHiggsCrossSectionWorkingGroup:2016ypw} which is in good agreement with the current measured values $\sigma(t\bar{t}H)_{\rm ATLAS} = 411^{+101}_{-92}$ fb \cite{ATLAS:2024gth} and $\sigma(t\bar{t}H)_{\rm CMS} = 466 \pm 96~({\rm stat.})^{+70}_{-53}~({\rm syst.})$ fb \cite{CMS:2020mpn}. The cross section of $t\bar{t}H$ becomes smaller in the case where $H$ is a pure pseudo-scalar due to phase space considerations. For $m_H = 120$ GeV, it was found that the $t\bar{t}H$ cross section for the case where $H$ is a ${\cal CP}$-odd scalar is a factor of $2.5$--$3.2$ smaller than the case where it is a pure ${\cal CP}$-even scalar depending on the center-of-mass energy \cite{Frederix:2011zi}. The sensitivity of the LHC on the ${\cal CP}$ phase depends on the decay branching channels of the top quark and the Higgs boson. For the pure ${\cal CP}$-even scalar, the number of the possible decay combinations is 42 while it is 30 for a ${\cal CP}$ odd scalar since $H \to WW^*$ and $H\to ZZ^*$ are not allowed. In Fig. \ref{fig:decays} we show the different decay branching ratios for $t\bar{t}H$ in the case of a pure ${\cal CP}$ even scalar (left) and a pure ${\cal CP}$ odd scalar (right). As expected, pure hadronic decays occur with the highest branching ratios but they are not efficient in the determination of the ${\cal CP}$ properties of the Higgs boson due to the huge associated QCD backgrounds and other effects such as pile-up. The number of efficient channels is therefore very small which implies that precisely measuring ${\cal CP}$ properties of the top-Higgs coupling at the LHC can be challenging.
 
\subsection{Observables for $\mathcal{CP}$ violation in the top–Higgs sector}

In addition to low-level and high-level variables that are indispensable for isolating $t\bar{t}H/tH$ signal from the overwhelming backgrounds at the LHC, the use of angular observables that are sensitive to the ${\cal CP}$ phase of the top-Higgs coupling is crucial. These observables have been studied extensively in the literature not only for the $t\bar{t}H$ production but for other new physics and SM processes as well (see {\it e.g.} Refs. \cite{Mahlon:1995zn,Gunion:1996xu,Gunion:1998hm,Field:2002gt,Valencia:2005cx,Mahlon:2010gw,Bernreuther:2013aga,Boudjema:2015nda,Shelton:2008nq,Godbole:2011vw, Rindani:2011pk,Prasath:2014mfa,Godbole:2015bda,Jueid:2018wnj,Arhrib:2018bxc, Godbole:2019erb,Ferroglia:2019qjy,Chatterjee:2019brg,Goncalves:2018agy,Cheung:2020ugr, Faroughy:2019ird}). These observables are found to be efficient in probing the top-Higgs coupling. The most important ones rely on the momentum of the charged lepton which has the highest spin-analyzing power ($\alpha_{\ell^\pm} = \mp 1$) \cite{Mahlon:1995zn}\footnote{While the down quark is a perfect spin analyzer at leading order, its identification at high-energy colliders is obstructed by the fact that light-flavor jets unlike those from heavy quarks (like charm or bottom jets) lack unique kinematic or hadronic features, making $d$-quark tagging effectively impossible in practice.}. It is worth mentioning  that constructing these observables in $t\bar{t}H$ production at the LHC is not easy due to several factors with the most important being the fact that many of them rely on the reconstruction of the $t\bar{t}H$ rest frame in the di-leptonic decays of the $t\bar{t}$ system. We classify these observables as follows:

\paragraph{Observables based on the $t/\bar{t}$ rest frame} These observables are based on full reconstruction of the $t/\bar{t}$ rest frame which is in principle challenging in the di-leptonic decay channels of the $t\bar{t}$ system. These observables were widely used in the study of spin-spin correlations or {\it entanglement} in $t\bar{t}$ production at the LHC; see Refs. \cite{ATLAS:2016bac,CMS:2019nrx} for recent measurements. The generic expression of the double differential cross section in $\cos\theta_{\ell^a} \cos\theta_{\ell^b}$ can be written as follows
\begin{equation}
   \frac{1}{\sigma} \frac{\text{d}^2\sigma}{\text{d}\cos\theta_{\ell^a} \text{d}\cos\theta_{\ell^b}} =
   \frac{1}{4}\bigg(1 + \alpha_{\ell^a} P_{a} \cos\theta_{\ell^a} + \alpha_{\ell^b} P_b \cos\theta_{\ell^b} + \alpha_{\ell^a} \alpha_{\ell^b} C_{ab} \cos\theta_{\ell^a} \cos\theta_{\ell^b}\bigg),
 \label{thetadouble}
\end{equation}
with $\alpha_{\ell^{a,b}} = \pm 1$ being the spin analyzing power of the charged lepton ($\ell^{a,b}$) and $\theta_{\ell^{a}} = \measuredangle (\hat{\ell}^{a}, \hat{S}_{a})$, $\hat{\ell}^{a}$ is the direction of flight of the charged lepton ($\ell^{a}$) in the $t/\bar{t}$ rest frame and $\hat{S}_{a}$ is the spin quantization axis in the basis $a$. There are commonly three bases for the spin quantization axis of the $t/\bar{t}$: ({\it i}) helicity basis where the quantization axis is defined as the direction of flight of $t/\bar{t}$ in the $t\bar{t}$ Zero-Momentum Frame (ZMF), ({\it ii}) transverse basis where the spin quantization axis is defined as the vector orthogonal to the plane composed by $t/\bar{t}$ direction of motion in the $t\bar{t}$ ZMF and the beam direction and ({\it iii}) $r$--basis in which the spin-quantization axis is defined to be orthogonal to the plane formed by the $t/\bar{t}$ direction of flight in the $t\bar{t}$ ZMF and the spin quantization axis in the transverse basis. 

\begin{figure}[!t]
    \centering
    \includegraphics[width=0.42\linewidth]{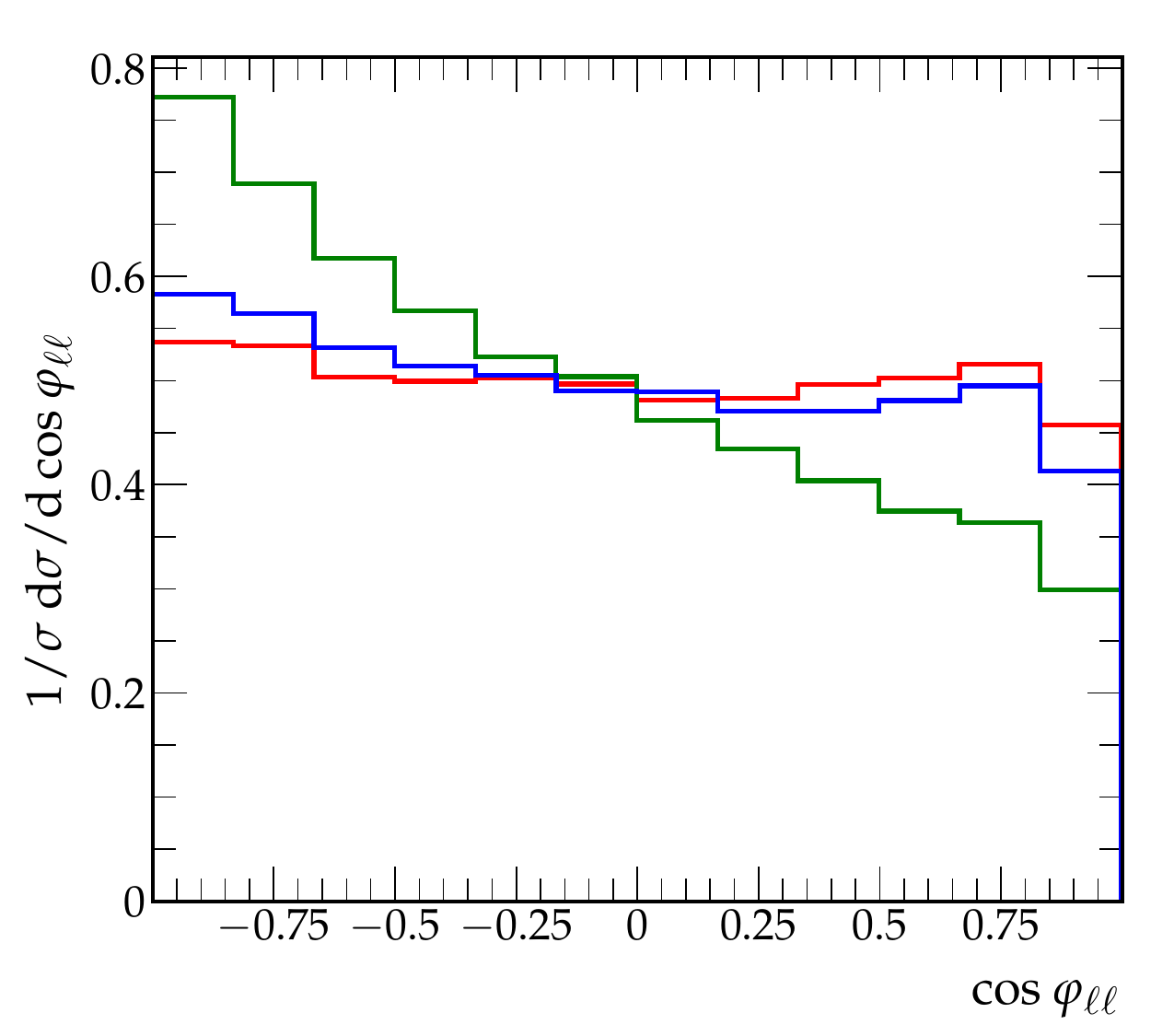}
    \includegraphics[width=0.42\linewidth]{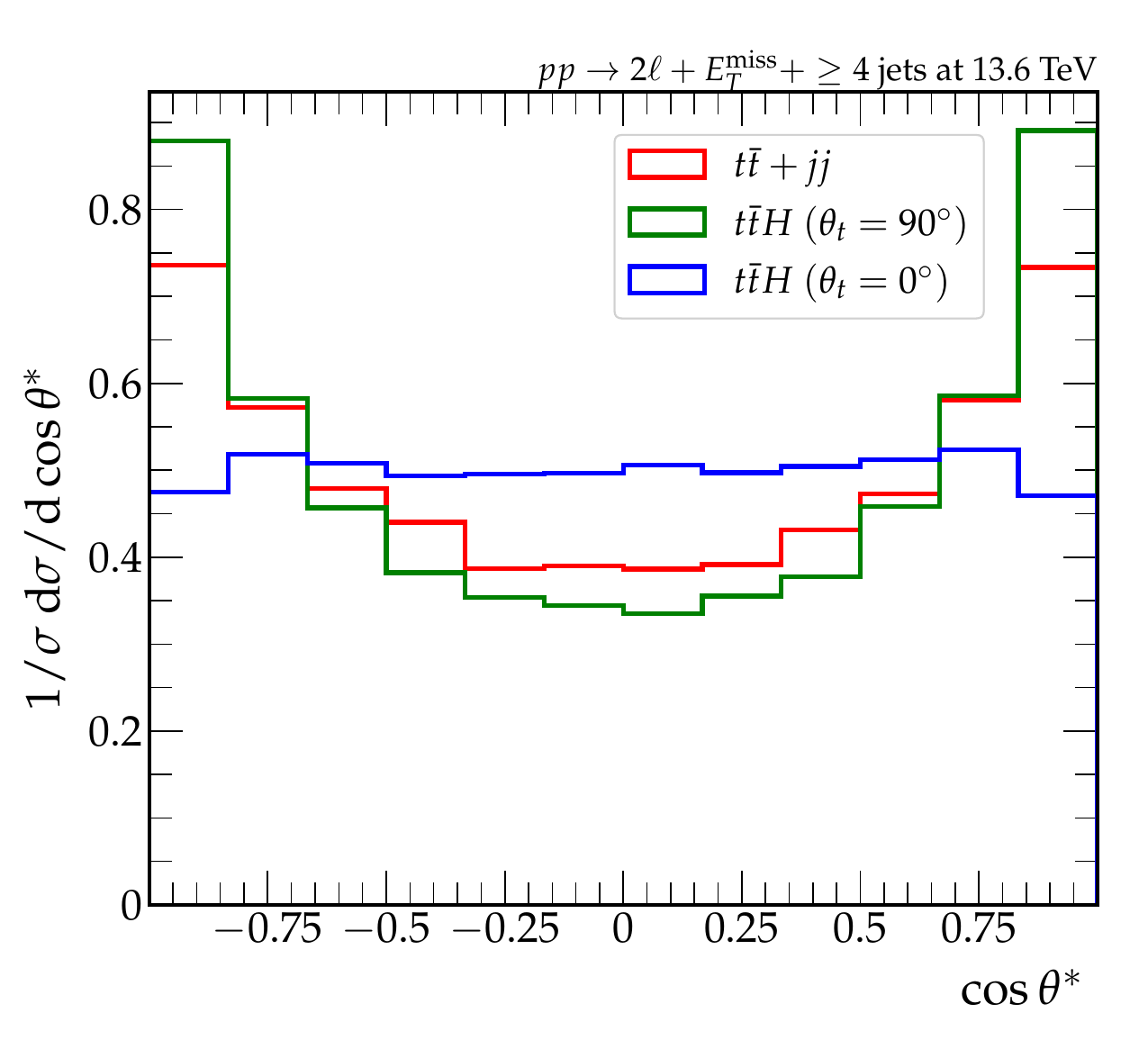}
    \caption{Comparison between $t\bar{t}+jj$ (red), $t\bar{t}H$ with $\theta_t = 90^\circ$ (green) and $t\bar{t}H$ with $\theta = 0^\circ$ (blue) for the cosine of $\varphi_{\ell^+\ell^-}$ defined in equation \ref{eq:varphill} (left) and the cosine of $\theta^*$ defined in equation \ref{eq:costhetaS} (right). These observables were constructed in the full di-leptonic decays of $t\bar{t}$ while assuming the Higgs boson to decay into $b\bar{b}$ at the detector level and $\sqrt{s} = 13.6$ TeV. The simplified detector simulation was performed using the \textsf{SFS} module \cite{Araz:2020lnp} of \textsf{MadAnalysis} \cite{Conte:2012fm,Conte:2014zja}. Theoretical predictions are based on the following simulation pipeline: \textsf{Madgraph\_aMC@NLO} \cite{Alwall:2014hca} for the matrix-element generation, \textsf{MadSpin} for the decays of $t/\bar{t}$ and the Higgs boson \cite{Artoisenet:2012st}, and \textsf{Pythia}~8.2 for parton showering and hadronization \cite{Sjostrand:2014zea}. The definition of jets, leptons and missing energy follows the same lines of Ref. \cite{Frank:2023epx}. Results are taken from Ref. \cite{Esmail:2024gdc}.}
    \label{fig:angles:MA5}
\end{figure}

\paragraph{Laboratory frame observables} They are defined in the $pp$ center-of-mass frame (also known as the laboratory frame). The most straightforward observable is the absolute value of the difference in the azimuthal angle of the two charged leptons coming from the dileptonic decays of the $t/\bar{t}$, {\it i.e.} 
\begin{eqnarray}
    \Delta \phi_{\ell^+ \ell^-}=|\phi_{\ell^+} - \phi_{\ell^-}|.
\end{eqnarray}
More complex observables were constructed in Ref. \cite{Boudjema:2015nda}. These are based on the full reconstruction of the Higgs boson candidate and they are found to have a good sensitivity in a study of ${\cal CP}$ violation in $t\bar{t}H$ at the parton level. The first observable is defined as
\begin{eqnarray}
\cos\theta_{\ell H} = \frac{(\hat{p}_{\ell^+} \times \hat{p}_{H}) \cdot  (\hat{p}_{\ell^-} \times \hat{p}_{H})}{|(\hat{p}_{\ell^+} \times \hat{p}_{H})|  |(\hat{p}_{\ell^-} \times \hat{p}_{H})|},
\label{eq:costhetaHL}
\end{eqnarray}
where $\hat{p}_{\ell^+}$, $\hat{p}_{\ell^-}$ and $\hat{p}_{H}$ are the directions of flight of $\ell^+$, $\ell^-$ and $H$ in the laboratory frame.  The sensitivity of this observable on the ${\cal CP}$ phase depends on the ability to keep track of the spin correlation between the Higgs boson and its decay products as it was found that for $H \to b\bar{b}$ this observable has a mild sensitivity \cite{Esmail:2024gdc}. Another angle can be formed from $\cos\theta_{\ell H}$ as 
\begin{eqnarray}
    \cos\widetilde{\theta}_{\ell H} = \lambda \cos\theta_{\ell H}, \quad \lambda = {\rm sign}((\hat{p}_{X} - \hat{p}_{\bar{X}})\cdot(\hat{p}_{\ell^-} \times \hat{p}_{\ell^+})),
\end{eqnarray}
where $X$ is the particle produced from the Higgs boson decay ($H \to X\bar{X}$). There are two other interesting angles that were constructed in Refs. \cite{Gunion:1996xu,Ferroglia:2019qjy}
\begin{eqnarray}
    b_2 = \frac{\left(\vec{p}_t \times \hat{p}_{\rm beam}\right)\cdot \left(\vec{p}_{\bar{t}} \times \hat{p}_{\rm beam}\right)}{|\vec{p}_t||\vec{p}_{\bar{t}}|}, \qquad b_4 = \frac{\left(\vec{p}_t \cdot \hat{p}_{\rm beam}\right)\cdot \left(\vec{p}_{\bar{t}} \cdot \hat{p}_{\rm beam}\right)}{|\vec{p}_t||\vec{p}_{\bar{t}}|},
    \label{eq:b2:b4}
\end{eqnarray}
with $\vec{p}_t$~($\vec{p}_{\bar{t}}$) being the three-momentum vector of $t$~($\bar{t}$) in the laboratory frame and $\hat{p}_{\rm beam}$ is the unit vector along beamline direction. Finally, we note that other observables based on the energies of the decay products of $t/\bar{t}$ were studied in Refs. \cite{Shelton:2008nq,Godbole:2011vw,Rindani:2011pk,Prasath:2014mfa,Godbole:2015bda,Jueid:2018wnj,Arhrib:2018bxc, Godbole:2019erb,Chatterjee:2019brg} but they have mild sensitivity on the ${\cal CP}$ phase of the top-Higgs coupling.

\paragraph{Observables based on the $t\bar{t}$ ZMF} These angles are build upon the direction of flight of $t/\bar{t}$, and charged leptons in the $t\bar{t}$ ZMF. The first angle is the opening angle between the two charged leptons in the $t\bar{t}$ ZMF defined as
\begin{eqnarray}
 \varphi_{\ell^+\ell^-} = \arccos\left({\frac{\hat{p}_{\ell^+} \cdot \hat{p}_{\ell^-}}{|\hat{p}_{\ell^+}| |\hat{p}_{\ell^-}|}}\right).
 \label{eq:varphill}
\end{eqnarray}
Another angle based on both the direction of flights of the $t/\bar{t}$ and $\ell^+,\ell^-$ is defined as \cite{Goncalves:2018agy} 
\begin{eqnarray}
    \Delta\phi_{\ell^+ \ell^-}^{t\bar{t}} = \tilde{\lambda} \arccos\bigg(\frac{(\hat{p}_{\ell^+}^{\rm ZMF} \times \hat{p}_{t}^{\rm ZMF}) \cdot (\hat{p}_{\ell^-}^{\rm ZMF} \times \hat{p}_{t}^{\rm ZMF})}{|\hat{p}_{\ell^+}^{\rm ZMF} \times \hat{p}_{t}^{\rm ZMF}| |\hat{p}_{\ell^-}^{\rm ZMF} \times \hat{p}_{t}^{\rm ZMF}|}\bigg), \quad {\rm with}~ \tilde{\lambda} = {\rm sign}[\hat{p}_t \cdot (\hat{p}_{\ell^+} \times \hat{p}_{\ell^-})]
\end{eqnarray}
while the second angle is more straightforward as defined as the one between the unit top momentum in the $t\bar{t}$ ZMF and the beamline direction, {\it i.e.}
\begin{eqnarray}
    \theta^* = \arccos\left({\frac{\hat{p}_{t}^{\rm ZMF} \cdot \hat{p}_{\rm beam}}{|\hat{p}_{t}^{\rm ZMF}| |\hat{p}_{\rm beam}|}}\right).
    \label{eq:costhetaS}
\end{eqnarray}

In Fig. \ref{fig:angles:MA5} we show the distributions of $\cos\varphi_{\ell^+\ell^-}$ and $\cos\theta^*$ for the three main processes: the irreducible QCD backgrounds ($t\bar{t}+jj$) and two signal hypotheses, {\it i.e.} $t\bar{t}H$ with $\theta_t=0^\circ$ and $t\bar{t}H$ with $\theta_t=90^\circ$. It is clear that these two angles are ${\cal CP}$--odd observables as they can distinguish between pure ${\cal CP}$ odd and pure ${\cal CP}$ even states. 

\begin{table}[!t]
\setlength\tabcolsep{5pt}
\renewcommand{\arraystretch}{1.1}
\begin{adjustbox}{max width=1.01\textwidth}
    \begin{tabular}{l l l l l}
    \toprule
    Analysis     &  Production mode & Final state & Limit & $L$ [fb$^{-1}$] \\
    \toprule
    ATLAS$_{7 \oplus 8}$     &  $t\bar{t}H~ \& ~tH$ & $H\to \gamma\gamma$  & $\kappa_t ~\in~[-1.3, 8.0]$ & $4.5 \oplus 20.3$~\cite{ATLAS:2014ayi} \\  
              &    &         $t \to {\rm all}$, $\bar{t} \to {\rm all}$ & & \\ \midrule
    ATLAS$_8$     & $t\bar{t}H$ & $H\to b\bar{b}$ & $\sigma/\sigma_{\rm SM} < 3.4$ & $20.3$~\cite{ATLAS:2015utn} \\ 
    & & $t/\bar{t} \to \geq 1 \ell$ + jets + $E_{T}^{\rm miss}$ &  \\ \midrule
    ATLAS$_8$     &  $t\bar{t}H$ & $H\to WW^*, ZZ^*, \tau^+\tau^-$ & $\sigma/\sigma_{\rm SM} < 4.7$ & $20.3$~\cite{ATLAS:2015xdt} \\ 
                  &    &         $t \to {\rm all}$, $\bar{t} \to {\rm all}$ & & \\ \midrule
    ATLAS$_8$     &  $t\bar{t}H$ & $H\to b\bar{b}$ & $\sigma/\sigma_{\rm SM} < 4.8$ & $20.3$~\cite{ATLAS:2016wki} \\
                      &    &         $t/\bar{t} \to 0\ell$ + jets & & \\ \midrule
    ATLAS$_{13}$     &  $t\bar{t}H$ & $H\to \gamma\gamma,VV^*,\tau\tau,b\bar{b}$ & $\sigma = 670 \pm 90~({\rm stat.})^{+110}_{-100}~({\rm syst.})$ fb & $79.8$~\cite{ATLAS:2018mme} \\
                      &    &         $t \to {\rm all}$, $\bar{t} \to {\rm all}$ & & \\ \midrule
    ATLAS$_{13}$    &  $t\bar{t}H~\&~tH$ & $H\to \gamma\gamma$ & $|\theta_t| < 43^\circ$ & $139$~\cite{ATLAS:2020ior} \\
        & & $t/\bar{t} \to \geq 1 \ell$ + jets + $E_{T}^{\rm miss}$ & $\mu = 1.43^{+0.33}_{-0.31}~({\rm stat.})^{+0.21}_{-0.15}~({\rm syst.})$ & \\ \midrule
    ATLAS$_{13}$     &  $t\bar{t}H$ & $H\to b\bar{b}$ & $\mu = 0.35^{+0.36}_{-0.35}$ & $139$~\cite{ATLAS:2021qou} \\ 
        & & $t/\bar{t} \to \geq 1 \ell$ + jets + $E_{T}^{\rm miss}$ &  \\ \midrule
    ATLAS$_{13}$     &  $t\bar{t}H~\&~tH$ & $H\to b\bar{b}$ & $\theta_t = 11^\circ\,{}^{+52^\circ}_{-73^\circ}$; ~ $\kappa_t = 0.84^{+0.30}_{-0.46}$ & $139$~\cite{ATLAS:2023cbt} \\ 
    & & $t/\bar{t} \to \geq 1 \ell$ + jets + $E_{T}^{\rm miss}$ &  \\ \midrule
    ATLAS$_{13}$    &  $t\bar{t}H$ & $H\to b\bar{b}$ & $\sigma = 411^{+101}_{-92}$ fb & $140$~\cite{ATLAS:2024gth} \\ 
    & & $t/\bar{t} \to \geq 1 \ell$ + jets + $E_{T}^{\rm miss}$ &  \\ \midrule
    CMS$_{7 \oplus 8}$     &  $t\bar{t}H$ & $H\to b\bar{b}$ & $\sigma/\sigma_{\rm SM} < 5.8$ &  $5.0 \oplus 5.1$~\cite{CMS:2013szn} \\ 
        & & $t/\bar{t} \to \geq 1 \ell$ + jets + $E_{T}^{\rm miss}$ &  \\ \midrule
    CMS$_{7 \oplus 8}$     &  $t\bar{t}H$ & $H\to {\rm hadrons, photons, leptons}$ & $\sigma/\sigma_{\rm SM} < 4.5$ & $5.1 \oplus 19.7$~\cite{CMS:2014tll} \\
            & & $t/\bar{t} \to \geq 1 \ell$ + jets + $E_{T}^{\rm miss}$ &  \\ \midrule
    CMS$_8$     &  $t\bar{t}H$ & $H\to b\bar{b}$ & $\sigma/\sigma_{\rm SM} < 4.2$ & $19.5$ \cite{CMS:2015enw} \\
            & & $t/\bar{t} \to \geq 1 \ell$ + jets + $E_{T}^{\rm miss}$ &  \\ \midrule
    CMS$_8$     &  $tH$ & $H\to b\bar{b},~\gamma\gamma,\tau\tau, WW^*$ & $\sigma/\sigma_{\kappa_t=-1} < 2.8$ &  $19.7$ \cite{CMS:2015nrd} \\
            &    &         $t \to {\rm all}$, $\bar{t} \to {\rm all}$ & & \\ \midrule
    CMS$_{13}$     &  $t\bar{t}H$ & $\ell + \tau_h + {\rm jets}$ & $\sigma/\sigma_{\rm SM} = 1.23^{+0.45}_{-0.43}$ & $35.9$         \cite{CMS:2018fdh} \\ \midrule
    CMS$_{13}$      &  $t\bar{t}H$ & $H \to b\bar{b}, gg$ & $\sigma/\sigma_{\rm SM} < 3.8$ & $35.9$    \cite{CMS:2018sah} \\
            &    &         $t/\bar{t} \to 0\ell$ + jets & & \\ \midrule
    CMS$_{7 \oplus 8 \oplus 13}$     &  $t\bar{t}H$ & $H \to b\bar{b},VV^*,\gamma\gamma,\tau\tau$ & $\sigma/\sigma_{\rm SM} = 1.26^{+0.31}_{-0.26}$ & $5.1 \oplus 19.7 \oplus 35.9$  \cite{CMS:2018uxb} \\ 
                  &    &         $t \to {\rm all}$, $\bar{t} \to {\rm all}$ & & \\ \midrule
    CMS$_{13}$     &  $t\bar{t}H$ & $H \to b\bar{b}$ & $\sigma/\sigma_{\rm SM} < 1.6$ & $35.9$ \cite{CMS:2018hnq} \\ 
            & & $t/\bar{t} \to  2 \ell$ + jets + $E_{T}^{\rm miss}$ &  \\ \midrule
    CMS$_{13}$     &  $tH$ & $H \to b\bar{b}, WW^*, \tau\tau,\gamma\gamma$ & $\kappa_t > -0.9$ & $35.9$  \cite{CMS:2018jeh} \\ 
                    &    &         $t/\bar{t} \to 2 \ell$ + jets + $E_T^{\rm miss}$ & & \\ \midrule
    CMS$_{13}$   &  $t\bar{t}H$ & $H \to \gamma\gamma$ & $|f^{Htt}_{\rm CP}| < 0.67$ & $137$  \cite{CMS:2020cga} \\
                  &    &         $t \to {\rm all}$, $\bar{t} \to {\rm all}$ & & \\ \midrule
    CMS$_{13}$   &  $tH ~\&~t\bar{t}H$ & $H \to \tau\tau, WW^*, ZZ^*$ & $\kappa_t~\in ]-0.9, -0.7[ \cup ]0.7, 1.1[$ & $137$~\cite{CMS:2020mpn} \\ 
                      &    &         $t \to {\rm all}$, $\bar{t} \to {\rm all}$ & & \\ \midrule
    CMS$_{13}$   &  $tH ~\&~t\bar{t}H$ & $H \to \tau\tau, WW^*$ & $C_t~\in [0.86, 1.26], \tilde{C}_t ~\in [-1.07, 1.07]$ & $140$~\cite{CMS:2022dbt} \\ 
                      &    &         $t \to {\rm all}$, $\bar{t} \to {\rm all}$ & & \\
    \toprule
    \end{tabular}
    \hspace{0.2cm}
    \end{adjustbox}
    \caption{Summary of the current sensitivities of ATLAS and CMS searches to the $\bar{t}tH$ analysis.  }
    \label{tab:ATLAS:searches}
\end{table}

\subsection{LHC bounds}
\label{sec:LHC}

The first constraints on $\kappa_t$ were obtained from measurements of the signal strengths in all the possible production and decay channels \cite{ATLAS:2016neq}. In that case, the information on $\kappa_t$ is extracted from the measurement of
\begin{eqnarray}
    \mu_{jj}^i \equiv \frac{\sigma(pp \to H)^i \times {\rm BR}(H \to jj)}{\sigma(pp \to H)^i_{\rm SM} \times {\rm BR}(H \to jj)_{\rm SM}},
\end{eqnarray}
$i$ refers to the production channels being included, {\it i.e.} ggF, VBF, $HZ$, $HW^\pm$ and $t\bar{t}H$ and $j$ to the decay channels such as $\gamma\gamma$, $gg$, $b\bar{b}$, $VV^*$ and $\tau^+ \tau^-$. The indirect measurement of $\kappa_t$ depends on the underlying assumption on the particles running in the loops of the Higgs decay into $\gamma\gamma$ and those in the loops of $gg\to H$. If one parametrizes the new physics contribution to $\Gamma(H)$ by $B_{\rm BSM}$, the early determination of $\kappa_t$ were found to be $\kappa_t = 0.87^{+0.15}_{-0.15}$ for $B_{\rm BSM} = 0$ and $\kappa_t = 1.43^{+0.23}_{-0.22}$ for $B_{\rm BSM} \neq 0$ \cite{ATLAS:2016neq}. Even though the recent measurements have improved the precision on $\kappa_t$ thanks to both the increase in the center-of-mass energy, and integrated luminosity, the results are strongly dependent on the underlying assumptions; particle running in the loops of $H\to gg$ and $H\to \gamma\gamma$, novel invisible decays such as those to light dark matter and undetected decays usually labeled as $\Gamma(H\to {\rm undetected})$ \cite{CMS:2018uag,ATLAS:2019nkf,CMS:2021kom}. Moreover it is very challenging to constrain the phase of the top-Higgs coupling using these methods since the observables being used are proportional to the square of the top-Higgs coupling, and thus called ${\cal CP}$-even observables. 

The search for direct $tH$ and $t\bar{t}H$ production at the LHC has started shortly after the discovery of the SM Higgs boson. During the Run~1 of the LHC at $7$ and $8$ TeV, both ATLAS and CMS aimed for establishing the presence of associated $tH/t\bar{t}H$ production using various decays of the Higgs boson including those to $WW^*, ZZ^*, \tau^+\tau^-, b\bar{b}$ and $\gamma\gamma$ and exploring all the possible $t/\bar{t}$ decays \cite{ATLAS:2014ayi,ATLAS:2015utn,ATLAS:2015xdt,ATLAS:2016wki,CMS:2013szn,CMS:2014tll,CMS:2015enw,CMS:2015nrd}. These analyses were not successful in observing direct production of $tH$ or $t\bar{t}H$ and they were use to place upper limits on their cross sections and constraints on the top-Higgs coupling modifier ($\kappa_t$). For instance the strongest bound on the cross section was set by ATLAS collaboration \cite{ATLAS:2015utn} using the Higgs decays into $b\bar{b}$ and exploring both the semi-leptonic and di-leptonic decays of the $t\bar{t}$; $\mu = \sigma_{t\bar{t}H}/\sigma_{t\bar{t}H}^{\rm SM} < 3.4$ at 95\% C.L. The strongest bound placed by the CMS collaboration on $\mu$ was $\mu < 4.2$ \cite{CMS:2015enw} assuming similar decay final states as the above-mentioned ATLAS analysis but with slightly lower luminosity. These early results were statistically limited and focused on establishing the rate of top-associated Higgs production, without sensitivity to possible CP-odd components in the top Yukawa coupling. At the LHC Run 2, both ATLAS \cite{ATLAS:2018mme} and CMS \cite{CMS:2018fdh} reported on the first observation of $t\bar{t}H$. The ATLAS collaboration reported the following value for the cross section; $\sigma(t\bar{t}H) = 670 \pm 90\,(\text{stat.})^{+110}_{-100}(\text{syst.})~\text{fb}$ while the first measurement of the signal strength by the CMS collaboration was $\mu = 1.23^{+0.45}_{-0.43}$ both in good agreement with the SM predictions at NLO-QCD.

These results have been significantly improved in further measurements \cite{ATLAS:2024gth,CMS:2020mpn} which marked a key transition from discovery to precision studies of the top-Higgs interaction. Following the observation of $t\bar{t}H$ production at the LHC, both ATLAS and CMS started dedicated analyses for the ${\cal CP}$ structure of the top-Higgs interaction. These analyses rely on the use of a few angular observables (such as $b_2$ and $b_4$ defined in equation \ref{eq:b2:b4}) and dedicated {\it multivariate techniques} such as BDTs (see next section for more details). Those variables in addition to other high-level kinematical variables were defined in several signal regions and are trained to distinguish between ${\cal CP}$-even, ${\cal CP}$-odd and mixed scenarios. First, ATLAS collaboration has performed two dedicated analyses for the study of the ${\cal CP}$ structure in the $H\to \gamma\gamma$ \cite{ATLAS:2020ior} and $H\to b\bar{b}$ final states \cite{ATLAS:2023cbt}. In the first analysis targeting the $H\to\gamma\gamma$ channel, basic kinematics and angular variables were used in the BDT training leading to $|\theta_t| < 43^\circ$ at the $95\%$ CL. While the second ATLAS analysis employed the $b_2$ and $b_4$ variables in addition to basic kinematic and angular variables and reported the following limit \cite{ATLAS:2023cbt}.
\begin{equation}
\theta_t = 11^\circ\,{}^{+52^\circ}_{-73^\circ}
\end{equation}
which are consistent with the SM pure ${\cal CP}$-even case. The study employing $H\to\gamma\gamma$ excludes the pure ${\cal CP}$-odd hypothesis by about 3.9$\sigma$. CMS collaboration has also studied the ${\cal CP}$ structure in the $H\to\gamma\gamma$ \cite{CMS:2020cga}, $H \to \tau^+\tau^-$ and $H \to WW^*$ \cite{CMS:2022dbt}. A BDT discriminant to distinguish between $t\bar{t}H$ and SM background is used followed by a further BDT discriminant to train ${\cal CP}$ odd versus ${\cal CP}$ even hypotheses using a ${\cal D}_0$ variable (see Ref. \cite{Gritsan:2016hjl} for more details about these observables). The results were translated into limits on $f_{\rm CP}^{Htt}$ defined as
\begin{eqnarray}
    f_{\rm CP}^{Htt} = \frac{|\tilde{C}_t|^2}{|\tilde{C}_t|^2 + |C_t|^2} \times {\rm sign}(\tilde{C}_t/C_t),
\end{eqnarray}
with $C_t = \kappa_t \cos\theta_t$ and $\tilde{C}_t = \kappa_t \sin\theta_t$. CMS reports a bound of $|f_{\text{CP}}^{Htt}| < 0.67$~\cite{CMS:2020cga}, which is insensitive to the sign of the CP-mixing angle because a key interference-sensitive observable was not included in the analysis (denoted as ${\cal D}_{\rm CP}$ in \cite{Gritsan:2016hjl}).

\section{Impact of Machine Learning}
\label{sec:new}
\subsection{Data format for machine learning analyses}

The first step in any problem that makes use of ML is defining the structure of the input data format. The performance of models depends heavily on how the data is structured. Different models, such as BDTs, MLPs, GNNs, and transformers, require distinct input formats.
BDTs work with structured, tabular data where each event is represented as a fixed-length feature vector. These features are typically high-level kinematic variables derived from raw particle data, including transverse momentum ($p_T$), pseudorapidity ($\eta$), invariant mass, and angular distributions. BDTs are well-suited for classification tasks with well-defined inputs and are widely used for event selection, background rejection, and high-level analysis. To probe the ${\cal CP}$ phase, angular distributions of final-state particles can be added to the input variables. These encode the ${\cal CP}$ structure of the top-Higgs coupling, and BDTs are often trained on such distributions to distinguish ${\cal CP}$-even from ${\cal CP}$-odd states. MLPs also take fixed-length feature vectors but can model more complex, nonlinear relationships between the input features. Unlike BDTs, which rely on decision trees, MLPs use deep learning (DL) to learn hierarchical feature representations, offering greater flexibility but requiring more data and tuning. GNNs operate on graphs where each particle is a node and edges represent relationships between the nodes $i$ and $j$ such as the lego distance; $\Delta R = \sqrt{(\eta_i - \eta_j)^2 + (\phi_i - \phi_j)^2}$  \cite{ExaTrkX:2020nyf,Abdughani:2018wrw,Moreno:2019bmu,Qasim:2019otl,Bernreuther:2020vhm,Shlomi:2020gdn,Alonso-Monsalve:2020nde,Esmail:2023axd,Guo:2020vvt,Dreyer:2020brq,Verma:2021ceh,Atkinson:2021nlt,Ma:2022bvt,Murnane:2023ksa,Konar:2023ptv}. In fully leptonic $\bar{t}tH$ decays, the graph has 11 nodes where each node corresponds to either $\ell^\pm$, $\nu, \bar{b}$, the four $b$ jets, the $t/\bar{t}$ quark and the Higgs boson. Node features include kinematic and angular properties; edge features capture distances. GNNs excel in jet tagging, event reconstruction, and particle tracking, and have been used to measure the ${\cal CP}$ phase \cite{Ren:2019xhp,Esmail:2024gdc}. Transformers, unlike MLPs or GNNs, process unordered sets of particles without explicit graph structures \cite{Araz:2024bom,Qu:2019gqs,Komiske:2018cqr,Dolan:2020qkr,Lee:2020qil,ATL-PHYS-PUB-2020-014,Onyisi:2022hdh}. They use self-attention to learn dependencies between particles. Each particle is represented independently, and attention layers learn their interactions. This “particle cloud” format allows transformers to analyze raw event data flexibly, while attention mechanisms highlight the most relevant particles.
The choice of data structure is critical. BDTs and MLPs are ideal when high-level features are well-defined. GNNs capture complex relations via graphs, and transformers provide scalability for raw particle data. Each method has trade-offs, and the best choice depends on the task. In the following, we discuss how each model is used to study the ${\cal CP}$ structure of the Higgs boson in the $t\bar{t}H$ channel.

\subsection{Machine learning methods for ${\cal CP}$ studies}

\subsubsection{BDT}

BDTs improve upon simple decision trees by reducing bias and variance through boosting. Using shallow trees helps prevent overfitting, while the ensemble captures complex patterns for a stronger classifier. Advantages of BDTs are numerous since they only require minimal preprocessing, handle both numerical and categorical data, and are robust to noise by correcting errors iteratively. However combining many tree increases the complexity of the BDTs. Moreover, although BDTs perform well on imbalanced datasets by emphasizing misclassified instances, their iterative nature leads to high computational cost and longer training times. Finally, hyperparameter tuning - particularly of the number of trees and the learning rate — is essential to prevent overfitting. The ATLAS and CMS collaborations used BDTs to probe the ${\cal CP}$ nature of the top–Higgs coupling exploiting various decay channels of the Higgs boson \cite{ATLAS:2020ior,ATLAS:2023cbt,CMS:2020cga,CMS:2022dbt}. As discussed in section \ref{sec:LHC}, these analyses simultaneously constrain $\kappa_t$ and $\theta_t$ through a binned profile likelihood fit using ${\cal CP}$-sensitive observables. To enhance the sensitivity, events passing a preselection are categorized in a two-step process. First, control and training regions are defined based on jet multiplicity, including those featuring $b$-tagged and large-$R$ jets. Training regions, which are signal enriched, are used to develop multivariate algorithms (MVAs)~\cite{TMVA:2007ngy}, where dedicated observables are constructed to probe ${\cal CP}$-violating effects. Due to the small expected contribution from $tH$ production, the categorization strategy, MVAs, and CP-sensitive observables are optimized for the $t\bar{t}H$ signal. Reconstruction BDTs are then trained to distinguish correct jet assignments from random ones by using relative kinematic information and invariant masses of jet pairs and triplets forming $W$-boson and $t/\bar{t}$-quark candidates. For each event, all possible jet permutations are evaluated, and the one with the highest BDT score is selected. These reconstruction BDTs not only improve the performance of the classification BDTs but also enable the extraction of observables sensitive to the ${\cal CP}$ structure of the top–Higgs coupling.

\subsubsection{MLP}

The MLP offers key advantages over BDTs in classification, particularly in modeling complex patterns within high-dimensional data. Unlike BDTs, which divide the feature space into discrete regions, MLPs use fully connected layers and non-linear activation functions to learn smooth, continuous decision boundaries, making them effective for correlated or high-dimensional features. Moreover, MLPs also generalize better across varied conditions. While BDTs need careful hyperparameter tuning to avoid overfitting, properly regularized MLPs can learn representations that generalize well. They are also more suitable for continuous parameter interpolation, as they capture functional dependencies between features and outputs, whereas BDTs rely on discrete splits that may miss smooth transitions. Recent studies have shown that MLPs can be parameterized over variables, enabling training on selected values and generalization across a full parameter range \cite{Baldi:2016fzo}. This is ideal for ${\cal CP}$ studies, where a network trained on specific ${\cal CP}$ phases can make robust predictions across the entire range. A conditional Deep Neural Network (DNN) extends this idea by incorporating extra input information, which are known as conditioning variables, that guide classification. These may require preprocessing, {\it e.g.} encoding or dimensionality reduction, and are integrated by concatenation or through dedicated layers. This allows the network focus on relevant features based on variables like the angle $\theta_{t}$, supporting interpolation across untrained $\theta_{t}$ values. Despite strong performance, MLPs can struggle with overlapping signal and background distributions. Applying initial selection cuts could improve signal-to-background ratios but may reduce effectiveness due to inter-correlated kinematic variables. Instead, de-correlation techniques, like square root of the covariance matrix or Gaussian transforms may be a solution to this problem \cite{TMVA:2007ngy}. However, it is more advisable not to apply these cuts to preserve the MLP flexibility in finding optimal decision boundaries.

\begin{figure}[!t]
    \centering
    \includegraphics[width=0.85\linewidth]{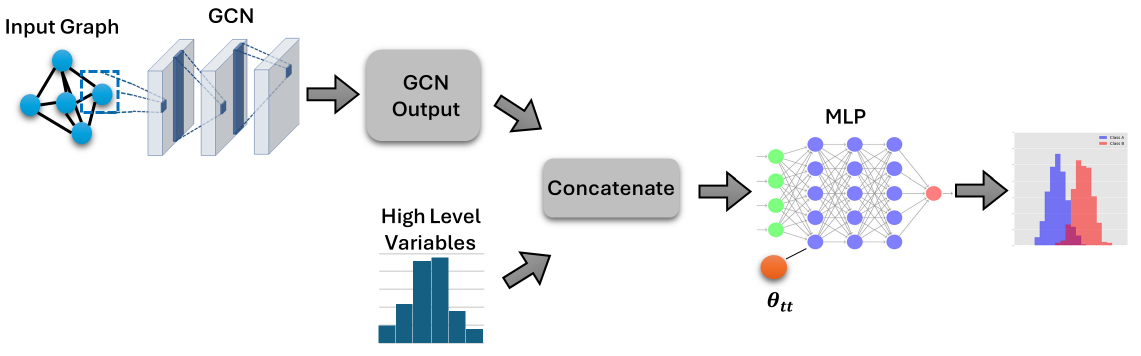}
    \caption{Schematic diagram of the multi-modal GNN illustrating the flow from the input graph through the GCN to produce an intermediate representation. This representation is then concatenated with high-level variables and passed through an MLP, resulting in the final output classification. Taken from \cite{Esmail:2024gdc}.}
    \label{fig:gnn_architecture}
\end{figure}

\subsubsection{GNN}
To construct a conditional GNN, a multi-modal architecture that integrates a Graph Convolutional Network (GCN) with an MLP is usually employed. The GCN component is particularly effective in capturing the topological relationships between nodes and edges, thereby enhancing the learning of graph-structured data. In this framework, reconstructed particles, their decay products, and parent particles are represented as nodes in the graph. 
As in~\cite{Esmail:2023axd}, each node $i$ has a feature vector 
\begin{equation*}
x = (I_1, I_2, I_3, \cdots, I_N, p_T, E, \eta, \phi),
\end{equation*}
with $p_T$, $E$, $\eta$ and $\phi$ being the basic low-level kinematic features and $I_1, \cdots, I_N$ are indicators that specify object types. In the case of ${\cal CP}$ studies in the $t\bar{t}H$, these indicators may refer to reconstrcuted objects such as {\it e.g.} leptons, $b$-jets, neutrinos, reconstructed $t/\bar{t}$-quarks and the Higgs boson, set to $1$ if present and to $0$ if not. GCNs are widely used for graph data due to their ability to learn topology-invariant representations~\cite{kipf2016semi,wu2019simplifying}. They generalize convolutions to graphs by averaging features from neighboring nodes, capturing both local structure and node attributes. The convolution of a graph $G \equiv (V, E)$ is given by 
\begin{equation}
    H^{(l+1)} = \sigma \left( \hat{D}^{-\frac{1}{2}} \hat{A} \hat{D}^{-\frac{1}{2}} H^{(l)} W^{(l)} \right),
\end{equation}
with $H^{(l)}$ being the feature matrix at layer $l$, $W^{(l)}$ the learnable weights, $\hat{A} = A + I_N$ the adjacency matrix with self-loops, and $\hat{D}$ the degree matrix. This operation updates node features through message passing. Traditional GCNs, however, rely on fixed adjacency matrices, limiting generalization to unseen graphs. They also treat all neighbors equally and use fixed aggregation schemes, reducing adaptability to diverse graph structures.
Dynamic Graph Convolutional Networks (DGCNNs)~\cite{manessi2020dynamic} overcome these issues with EdgeConv, which focuses on edge-based information flow. DGCNNs dynamically learn edge weights during training, adapting to graph structure and task demands. Rather than relying on static metrics like the lego distance $\Delta R$, edge weights are learned directly. EdgeConv is defined as
\begin{equation}
    \mathbf{h}_{ij} = \Phi(\mathbf{v}_i, \mathbf{v}_j - \mathbf{v}_i),
\end{equation}
where $\mathbf{v}_i$ is the feature vector of node $i$, and $\Phi$ is a shared MLP. The updated feature of node $i$ is:
\begin{equation}
    \mathbf{v}_i' = \rho(\{\mathbf{h}_{ij} \mid j \in N(i)\}),
\end{equation}
where $\rho$ is a symmetric aggregation function like for example the maximum or average pooling. Edge features are recalculated at each layer, letting the graph structure evolve dynamically with the learned features EdgeConv key advantage. An example of a GCN diagram is shown in Fig. \ref{fig:gnn_architecture}.

\subsubsection{DL-based Likelihood Inference}%
In addition to the conditional networks, DL-based likelihood inference can be employed to test the ${\cal CP}$ phase using Monte Carlo (MC) simulated events. In LHC analyses, a likelihood function is typically constructed as $p(x_i \mid \theta)$, representing the probability of observing a set of events characterized by observables $x_i$, given a model parameter $\theta$. In the context of ${\cal CP}$ studies, this likelihood function can be adapted by defining $x_i$ in terms of kinematic and angular observables, while $\theta$ corresponds to the ${\cal CP}$ phase value ($\theta_t$). The full likelihood function is modeled as a Poisson distribution, as formulated in Ref. \cite{Bahl:2021dnc}.
\begin{equation}
p_{\rm full}(x_i \mid \theta) = \rm Pois \left(n\mid L\sigma(\theta_t)\right) \prod_i p(x_i\mid \theta_t)    \,,
\end{equation}
with $n$ being the number of events, $L$ is the integrated luminosity and $\sigma(\theta_t)$  is the total cross section as a function of $\theta_t$. The function $p(x_i\mid\theta)$ represents the probability of observing a variable $x$ given the model parameter $\theta_t$  using MC simulations. The observed variables, $x_i$, often estimated using MC simulation then add extra physical effects such as parton showering, hadronization, hadron decays and detector effects. The effects of these lead to  
\begin{equation}
    p(x\mid\theta) = \int dz_{\rm d}\int dz_{\rm s} \int dz_{\rm MC} \ p(x\mid z_{\rm d})  p(z_{\rm d}| z_{\rm s})  p(z_{\rm s}\mid z_{\rm MC}) p(z_{\rm MC}\mid \theta)\,,
\end{equation}
with the latent variables $z_d,z_s,z_{\rm MC}$ represent the detector, parton shower and parton level observables, respectively.  A problem arises due to the large number of the latent variables in the integrals. This can be alleviated by constructing the likelihood ratio 
using two model parameters, $\theta_1$ and $\theta_2$ \cite{Brehmer:2018eca}
\begin{equation}
    r(x,z\mid \theta_1,\theta_2) \equiv \frac{p(x,z\mid \theta_1)}{p(x,z\mid \theta_2)}=\frac{d\sigma(z_{\rm MC}\theta_1)\sigma(\theta_1)}{d\sigma(z_{\rm MC}\theta_2)\sigma(\theta_1)}\,,
\end{equation}
with $d\sigma(z_{\rm MC}\mid \theta_i)$ being represents the parton level event weights. The second independent parameter being $\kappa_t$ and $\sigma(\theta_2)$ is the total cross section as a function of $\kappa_t$. Noticeably, when the likelihood ratio is considered, the integration over the parton shower and detector observables cancel out ending up with three dependent parameters, $\theta_1$, $\theta_2$ and the MC simulation. Similarly, a joint likelihood ratio can be constructed from the parton level events weights as \cite{Brehmer:2019xox}
\begin{equation}
    t(x,z\mid\theta) \equiv \nabla_\theta \log p(x,z\mid\theta) = \frac{\nabla_\theta d\sigma(z_{\rm MC}\mid\theta)}{d\sigma(z_{\rm MC}\mid\theta)} -\frac{\nabla_\theta\sigma(\theta)}{\sigma(\theta)}\,.
\end{equation}
The joint likelihood ratio can evaluated from the derivatives of the weights of the parton-level events and the  total cross section for a given theoretical parameter. \textsf{MadMiner} implements a morphing technique to efficiently evaluate event weights across the entire parameter space of the model. This approach facilitates the computation of the joint likelihood ratio and joint scores for all MC simulated events. The joint likelihood ratio  can be used to define suitable loss functions whose minimizing function is the true likelihood ratio $p(x\mid\theta_1,\theta_2)$ \cite{Brehmer:2018hga,Brehmer:2018kdj}
\begin{equation}
    \mathcal{L} = \frac{1}{N} \sum_i \left|r(x_i,z_i\mid\theta_1,\theta_2)- \hat{r}(x_i\mid\theta_1,\theta_2)\right|^2  \,,
\end{equation}
with $N$ is the sample size, and the sum runs over events in the sample. Minimization of the loss function can be obtained using a neural network, {\it i.e.} MLP, with the standard training techniques. In this case, the network tries to evolve the estimated function  $\hat{r}(x_i\mid\theta_1,\theta_2)$ from the true MC sampled one  $r(x_i,z_i\mid\theta_1,\theta_2)$. 

\subsection{Summary of the expected limits}

In this subsection, we summarize the expected limits on the ${\cal CP}$ phase ($\theta_t$) and the coupling modifier ($\kappa_t$) in studies that made use of MLP, GCN, and \textsf{MadMinder}. The results obtained with BDTs are summarized in section \ref{sec:LHC}.

\paragraph{MLP} As shown in \cite{Esmail:2024gdc}, leveraging high-level kinematic distributions and employing an MLP model enhance the distinction between signal and background events . These distributions encode essential information about the overall event structure, allowing the MLP, with its fully connected layers, to effectively capture global patterns and improve classification performance. Additionally, the network architecture integrates a linear layer that encodes the conditional parameter $\theta_{t}$, enabling the MLP to learn relevant features that enhance the signal-to-background ratio for each value of $\theta_{t}$. This setup allows the model to interpolate across values of $\theta_{t}$ that were not explicitly included in training, extending its sensitivity beyond the trained parameter space. The conditional MLP is trained on seven ${\cal CP}$ phase values with $\theta_t = 0^\circ,\pm 30^\circ,\pm 45^\circ,\pm 90^\circ$ and tested on all the other ${\cal CP}$ phase values. At the HL-LHC and for the full-leptonic decays of the $t\bar{t}$ system and Higgs boson decaying into $b\bar{b}$, it is found that the MLP can constrain ${\cal CP}$ angles of
\begin{eqnarray}
    \theta_t \in ~ [-60^\circ, 60^\circ]~([-10^\circ, 10^\circ]),\qquad {\rm for}~~ {\cal L} = 300~(3000)~{\rm fb}^{-1},
\end{eqnarray}
assuming $\kappa_t = 1$.

\paragraph{GNN} In \cite{Esmail:2024gdc}, DGCNN was incorporated with an MLP, for high level features encoding, and a linear layer to parametrize the ${\cal CP}$ phase. The network is trained on ${\cal CP}$ phase values of $\theta_t = 0^\circ,\pm 30^\circ,\pm45^\circ,\pm 90^\circ$ and tested on all ${\cal CP}$ phase values. At the HL-LHC and for $\kappa_t = 1$, ${\cal CP}$ phases of $[-45^\circ, 30^\circ]$~($[-5^\circ, 5^\circ]$) are constrained assuming $300$~($3000$) fb$^{-1}$ of integrated luminosity.

\paragraph{\textsf{MadMiner}} In Ref. \cite{Barman:2021yfh}, studies of the ${\cal CP}$ properties of the top-Higgs coupling in $t\bar{t}H$ where the Higgs boson decays into $\gamma\gamma$. The input features to \textsf{MadMiner} include $80$ reconstructed kinematic and polarization variables, as proposed in the ATLAS Higgs-top CP study \cite{ATLAS:2020ior}. By combining semi-leptonic, di-leptonic and hadronic decays of the $t\bar{t}$ system, they reported the following expected bounds $|\kappa_t|\lesssim 8\%$ and $ |\theta_t| \lesssim 13^\circ$ at the $68\%$ confidence level at the HL-LHC.

\section{Going beyond} 
\label{sec:prospects}

In this section, we summarize two promising algorithms, transformers and heterogeneous graph networks, which can be employed in the studies of ${\cal CP}$ properties of the top-Higgs coupling in $t\bar{t}H/tH$ at the LHC.

\subsection{Transformers}
An emerging direction in ${\cal CP}$ studies is the application of transformer networks. Originally developed for Natural Language Processing (NLP), transformers are now widely used in particle physics due to their ability to model high-dimensional and non-ordered data with long-range dependencies. Using self-attention, they dynamically assess the relevance of each input element well suited for collider events where interactions are complex and non-local. Transformers have shown strong performance in jet tagging~\cite{Qu:2022mxj,DiBello:2022iwf,Tagami:2024gtc,Brehmer:2024yqw,Wu:2024thh,Blekman:2024wyf,He:2023cfc,Finke:2023veq,Qiu:2022xvr,Tomiya:2023jdy,Spinner:2024hjm,Hammad:2024qme,Hammad:2024cae}, event classification~\cite{Hammad:2023sbd,Hammad:2024hhm,Mikuni:2021pou,Gao:2024zdz,Builtjes:2022usj}, MC event generation~\cite{Kach:2022uzq}, and track reconstruction~\cite{VanStroud:2024fau}. Their set-based nature avoids fixed input structures, enabling them to capture subtle correlations beyond traditional models and they also support generative modeling which offers data-driven alternatives to classical MC event generation. A key advantage is their scalability and interpretability since attention maps offer insight into decision-making by identifying which particles or features drive classification. This is may be especially valuable for ${\cal CP}$ studies of the top-Higgs coupling, where interpretability aids in distinguishing signal from backgrounds. As computational techniques advance, transformer-based architectures are expected to play an increasingly central role in data-driven high-energy physics. Transformer networks belong to a class of models that naturally exhibit permutation invariance \cite{guo2021pct} is enabled by attention mechanisms that weigh input features based on relative importance. 
Mathematically, given an input $X_{i,j}$ and a second set $S_{n,m}$, the queries ($Q$), keys ($K$), and values ($V$) are given by:
\begin{equation}
    Q^{i \times j} = X W_Q,\quad K^{n \times j} = S W_K,\quad V^{n \times j} = S W_V.
\end{equation}
The attention weights $\alpha$ are defined from $Q$ and $K$ as
\begin{equation}
    \alpha^{i \times n} = \text{softmax}\left(\frac{Q K^T}{\sqrt{d}} \right),
\end{equation}
with output being
\begin{equation}
    \mathcal{Z}^{i \times j} = \alpha V.
\end{equation}
Multi-head attention runs multiple attention operations in parallel, each producing an output $\mathcal{Z}_{i \times j}$. The final output is thus
\begin{equation}
    \mathcal{O}^{i \times j} = \text{concat}(\mathcal{Z}_1, \dots, \mathcal{Z}_n) W,
\end{equation}
which is followed by a residual connection $\widetilde{X}^{i \times j} = X + \mathcal{O}$.
This produces expressive representations that capture dependencies among particles. Despite their success in other physics applications, transformers have not yet been used for studying the ${\cal CP}$ nature of the top-Higgs coupling, largely due to challenges in incorporating ${\cal CP}$-sensitive features. One solution is to embed ${\cal CP}$-sensitive variables as input tokens; like the one that was employed in Ref. \cite{Hammad:2024hhm}. Another approach to use transformers in ${\cal CP}$ studies may benefit from introducing pairwise observables as a bias in the attention score, {\it i.e.}
\begin{equation}
    \text{Attention} = \text{softmax}\left(\frac{Q K^T}{\sqrt{d}} + U\right) \cdot V,
\end{equation}
where $U$ encodes ${\cal CP}$-sensitive interactions~\cite{Qu:2022mxj,Wu:2024thh}. These strategies enable transformers to learn subtle ${\cal CP}$-violating patterns in collider data.

\subsection{Heterogeneous Graphs}

One promising approach involves exploiting structured relationships among network inputs to boost discrimination between signal and backgrounds. A viable strategy is in constructing heterogeneous graphs as inputs to Graph Neural Networks (GNNs) \cite{Esmail:2024jdg,Huang:2023ssr}. These graphs connect final-state particles with irregular yet meaningful structure, allowing for customized input representations that enhance the GNN's sensitivity to the ${\cal CP}$ phase. Another possible improvement beyond the current proposed DL methods for ${\cal CP}$ analysis is to construct Heterogeneous graphs from the final state particles.  Heterogeneous graphs offer significant advantages over fully connected graphs, particularly in scenarios where complex relationships among diverse entities need to be captured. While fully connected graphs ensure that every node is linked to every other node, making information readily accessible across the entire structure, this approach can lead to inefficiencies, increased computational complexity, and difficulties in distinguishing between relevant and irrelevant connections. In contrast, heterogeneous graphs introduce a more structured and meaningful way of representing relationships by distinguishing between different types of nodes and edges.
One of the primary advantages of heterogeneous graphs is their ability to encode richer structural information. In a fully connected graph, all nodes are treated equally, and every node has a direct link to all others, which often results in redundant or noisy connections. Insufficient detail like this can obscure meaningful relationships and lead to inefficient learning, as the model must process many unnecessary interactions. Heterogeneous graphs, on the other hand, allow for the explicit modeling of different entity types and their specific interactions. For instance, in a collider physics application, a heterogeneous graph can differentiate between particle types, interactions, and detector components, preserving the underlying physical relationships rather than forcing an all-to-all connectivity.
Another critical improvement offered by heterogeneous graphs is their ability to reduce computational overhead. Fully connected graphs become computationally expensive as the number of nodes increases, requiring storage and processing of a large adjacency matrix. The number of edges in a fully connected graph scales quadratically with the number of nodes, which can lead to infeasibility in large-scale applications. Heterogeneous graphs, however, maintain a sparse and structured connectivity, where only meaningful edges are retained based on domain knowledge or learned representations. This sparsity significantly reduces memory consumption and accelerates computation while preserving essential interactions. Heterogeneous graphs also enhance interpretability by incorporating domain-specific priors. In a fully connected graph, all connections are treated equally, making it difficult to discern which relationships are crucial for decision-making. However, by explicitly defining different types of nodes and edges, a heterogeneous graph can represent domain-specific relationships, enabling the model to make more informed and interpretable decisions. In \cite{Esmail:2024jdg}, heterogeneous graphs were used to study the ${\cal CP}$ nature of the Higgs boson produced in gluon-gluon fusion and decaying into a pair of $\tau$ leptons. The great performance of heterogeneous graphs in that study is encouraging to pursue similar analyses of ${\cal CP}$ phase of the top-Higgs coupling in $t\bar{t}H$ production at the LHC.

\section{Summary}
\label{sec:summary}
DL represents the cutting edge of research in particle physics and other disciplines. Despite its success, training and evaluating DL models remains a resource intensive task, both in terms of time and computational power. To mitigate these challenges, the conditional computation framework has been proposed. This method enhances efficiency by activating only selected components of a neural network during each forward pass, rather than utilizing the entire architecture.
In this paper, we  review the recent progress in the analyses of the ${\cal CP}$ structure of the top-Higgs coupling  at the LHC using DL methods. The measurement of both the absolute magnitude ($\kappa_t$) and the phase ($\theta_t$) of the top-Higgs coupling is one of the important Higgs programs at the LHC as these will provide invaluable information for searching new physics models beyond the SM especially those influencing the mechanism of electroweak symmetry breaking and those providing solutions to the problem of baryon asymmetry in the universe. While current LHC limits reported on by the ATLAS and CMS collaborations primarily rely on BDTs, more advanced approaches — such as those utilizing DGCNNs and transformers — have the potential to significantly improve the current limits. In fact, some phenomenological studies have found that ${\cal CP}$ phases of order $5^\circ$--$10^\circ$ can be constrained if methods like conditional DGCNNs or MLPs are used. Transformers which were found to be very successful in various particle-physics analyses can further enhance the sensitivity of the LHC to the top-Higgs coupling.

\section*{Acknowledgements}
AH is funded by the Grant Number 22H05113 from the ``Foundation of Machine Learning Physics'', the ``Grant in Aid for Transformative Research Areas''  22K03626 and the Grant-in-Aid for Scientific Research (C). The work of AJ is supported by the Institute for Basic Science (IBS) under Project Code IBS-R018-D1.

\bibliographystyle{JHEP}
\bibliography{main.bib}

\end{document}